\documentclass[twoside]{ilcws10}
\usepackage[latin1]{inputenc}
\usepackage[dvips]{graphicx,epsfig,color}
\usepackage{wrapfig,rotating}
\usepackage{amssymb,amsmath,array}

\usepackage[sort&compress,square,numbers]{natbib}
\usepackage{hyperref}

\begin{document}
\title{
Simulation of Beam-Beam Background at CLIC} 
\author{André Sailer
\vspace{.3cm}\\
CERN - PH-LCD \\
Geneva  - Switzerland
}
%
%
\maketitle

\begin{abstract}
  The dense beams used at CLIC to achieve a high luminosity will cause a large
  amount of background particles through beam-beam interactions. Generator level
  studies with \textsc{GuineaPig} and full detector simulation studies with an
  ILD based CLIC detector have been performed to evaluate the amount of
  beam-beam background hitting the vertex detector.
\end{abstract}

\section{Introduction}
The Compact Linear Collider (CLIC) is designed for electron-positron collisions
at c.m. energies up to 3~TeV~\cite{braun2008}. The normal-conducting RF cavities
operate at a gradient of 100~MV/m, and the RF power is distributed along the
accelerator by a 2.4~GeV, high intensity "drive beam". In order to achieve the
desired luminosity, CLIC operates with bunch trains of 312~bunches separated by
0.5~ns, with a repetition rate of 50~Hz and transverse bunch size of
$45~\mathrm{nm}\mathrm{~by~}1~\mathrm{nm}$ and a length of $44~\mathrm{\mu m}$. A
number of design considerations have led to a crossing angle at the interaction
point of 20~mrad. The large number of electron-positron pairs produced by the
beam-beam interaction has to be studied in a full detector simulation to
evaluate and minimize the impact on the detector performance.

\section{Detector simulation and forward region}


\begin{figure}[b]\centering
  \centerline{\includegraphics[width=0.95\columnwidth, clip]{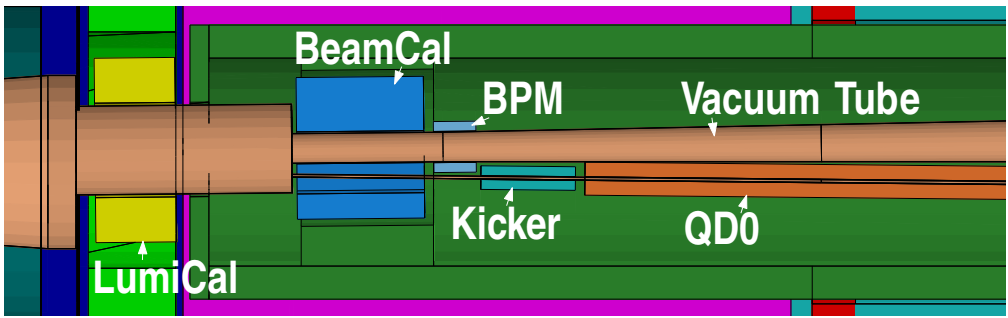}}
  \caption{Forward region layout used in the detector simulation.}\label{fig:fwxz}
\end{figure}

The detector used for the simulation is based on the ILD detector
concept~\cite{ILD:2009} for the ILC. A few modifications are needed to adapt the
detector for CLIC: Most importantly, a 4~Tesla solenoid field without Anti-DID
is used, because the Anti-DID reduces the luminosity at CLIC by about
20\%~\cite{dalena2010}. The vertex detector (VXD) consists of three double
layers, each with a total length of 25~cm and a radius of 31, 46 and 60 mm. The
inner most layer is placed at twice the radius compared to the VXD of the
ILC-ILD.

The forward region of a detector at CLIC (Figure \ref{fig:fwxz}) has to provide
the same functionality as for a detector at the ILC. The goal is to keep the
number of particles back-scattering into the central detectors as small as
possible while still maintaining an angular coverage down to polar angles of a
few millirad if possible. The CLIC forward region therefore contains most of the
elements that are planned for the ILC: A Luminosity Calorimeter (LumiCal) to
precisely count the number of Bhabha events in an angular region between about
40 and 100 mrad to measure the luminosity; a Beam Calorimeter (BeamCal),
extending the angular coverage of the forward calorimeters down to polar angles
of about 10~mrad. The forward region also contains masking to keep particles
produced by the beam-beam interaction from back-scattering into the main
detectors and to protect the equipment downstream of BeamCal, such as the beam
position monitor (BPM) and kicker of the intra train feedback, and the final
focus quadrupole (QD0).

Because of the small radius of the VXD and the low energetic nature of the
particles back-scattering from the forward region, a large amount of hits in the
VXD can be expected from this background. The hit density from beam-beam
background will be presented as an example for its impact on the detector.



\section{Simulation of beam-beam effects at CLIC}
\begin{wrapfigure}{O}{0.5\columnwidth}
  \centerline{\includegraphics[width=0.45\columnwidth]{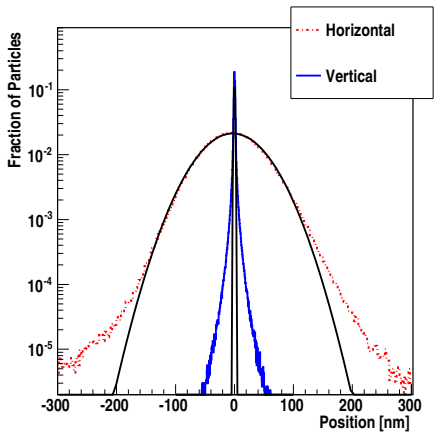}}
  \caption{Beam profiles used for the simulation and Gaussian fits to the profiles.}\label{fig:beamprof}
\end{wrapfigure}

The beam-beam interactions and background pair production has been simulated
with \textsc{Gui\-nea\-Pig}~\cite{schulte1996,schulte1998,gpcode}. Because the
charge distribution of the beams coming from the CLIC accelerator and beam
delivery system can not be described by a simple approximation (Figure
\ref{fig:beamprof}), files containing the proper distribution of particles in
the bunches have to be used as an input. A Gaussian approximation to the charge
distributions leads to a different amount of background particles. To produce
uncorrelated bunch crossings, 312 files each for the electron and positron
bunches were used, corresponding to a full CLIC bunch train~\cite{gpfiles}.

\begin{figure}[tp]
  \begin{tabular}{l r}
    \includegraphics[width=0.45\columnwidth]{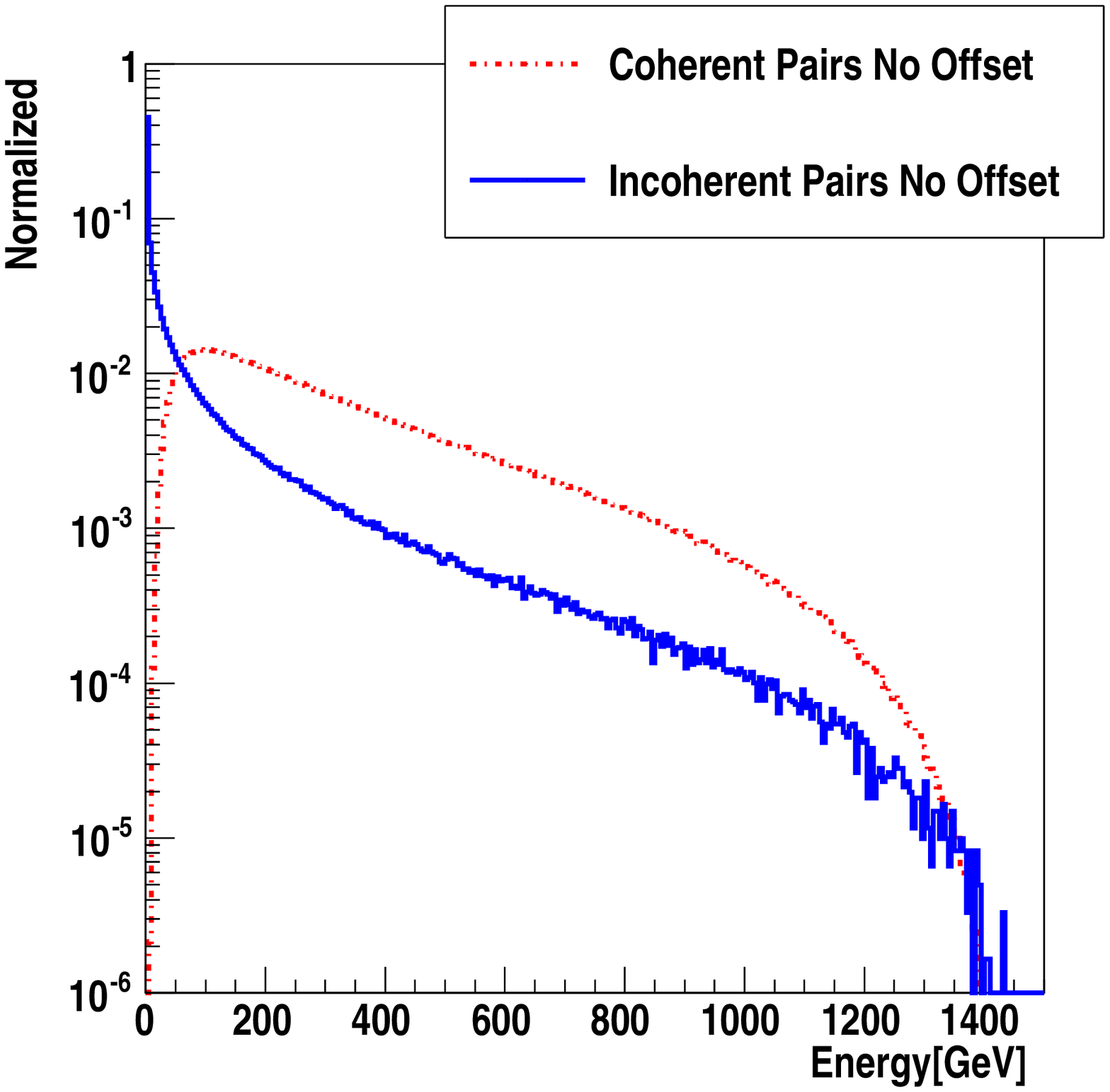} &
    \includegraphics[width=0.45\columnwidth]{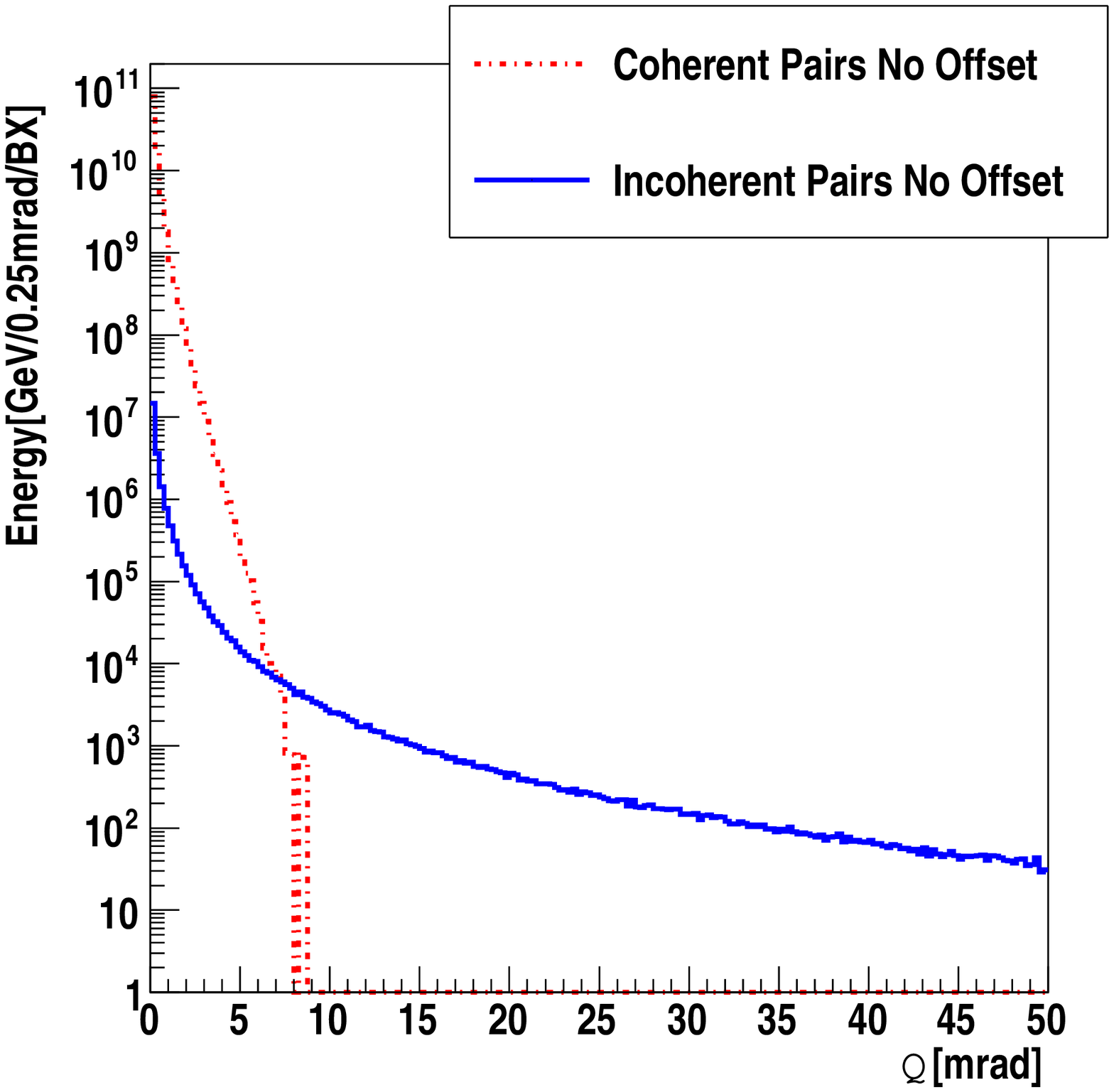}
  \end{tabular}
  \caption{Left: Energy spectrum of the background pairs. Right: Distribution of
    the energy of the coherent and incoherent pairs per production angle after
    deflection.}\label{fig:energyangle00}
\end{figure}
During one bunch crossing at CLIC about $3.1\cdot 10^5$ particles from
incoherent pairs are produced with an energy above 5~MeV. They are called
incoherent pairs, because the pair is produced, when real or virtual photons
interact with an electron~\cite{schulte1996}. An electron-positron pair can also
be produced, when a photon interacts with the coherent field of the bunch. At
CLIC $3.3\cdot10^8$ of these so called coherent pairs are
produced~\cite{dalena2010_2}. Because of their higher cut-off energy (Figure
\ref{fig:energyangle00} left) the coherent pairs are deflected less by the
oncoming bunch and leave the detector without interaction if the outgoing beam
pipe and hole in BeamCal are large enough, i.e.  approximately 10~mrad (Figure
\ref{fig:energyangle00} right).

\section{Imperfect beam collisions and background}
\label{sec:nonom}
\begin{figure}[tpb]
  \begin{tabular}{l r}
    \includegraphics[width=0.45\columnwidth]{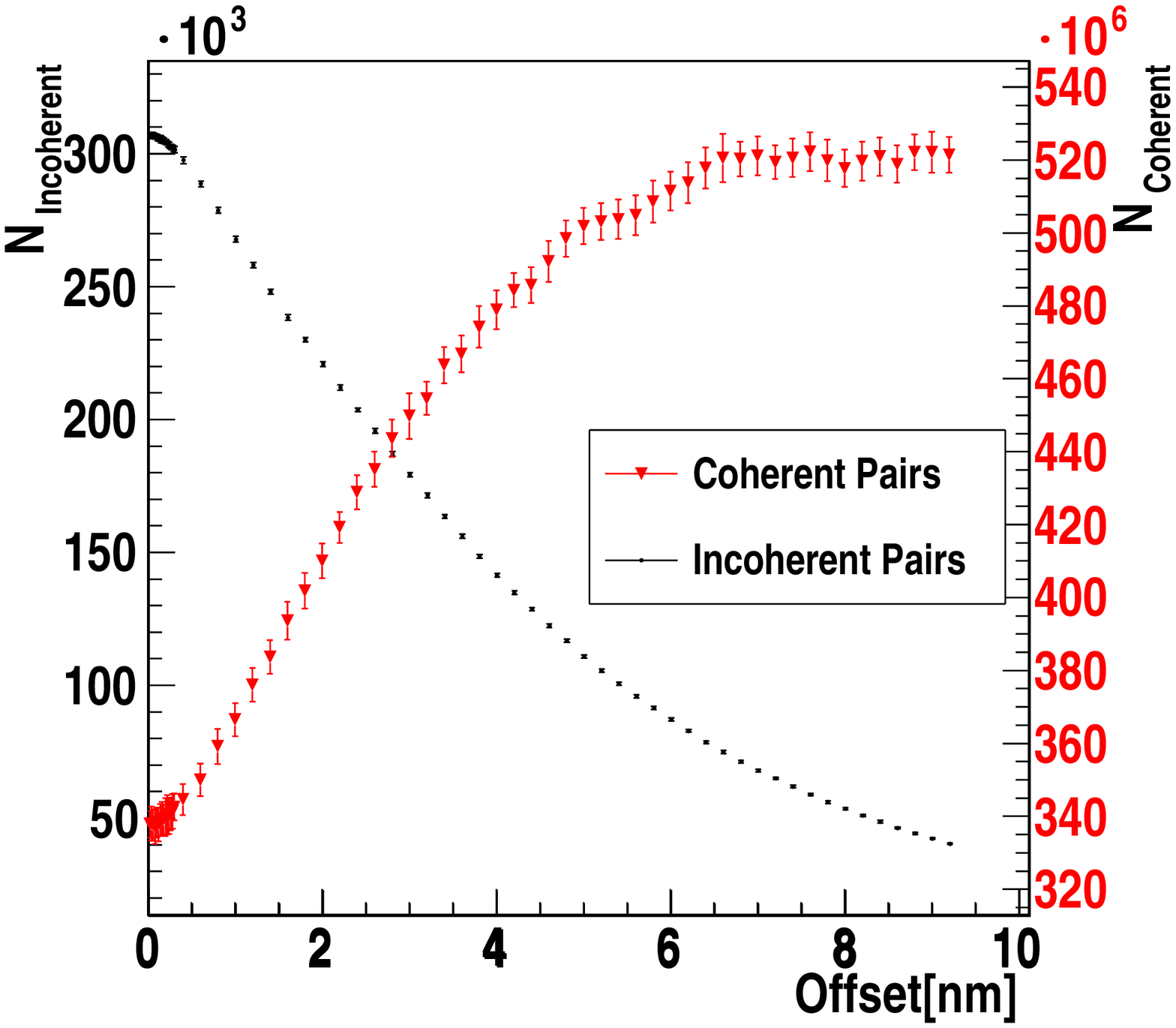} &
    \includegraphics[width=0.45\columnwidth]{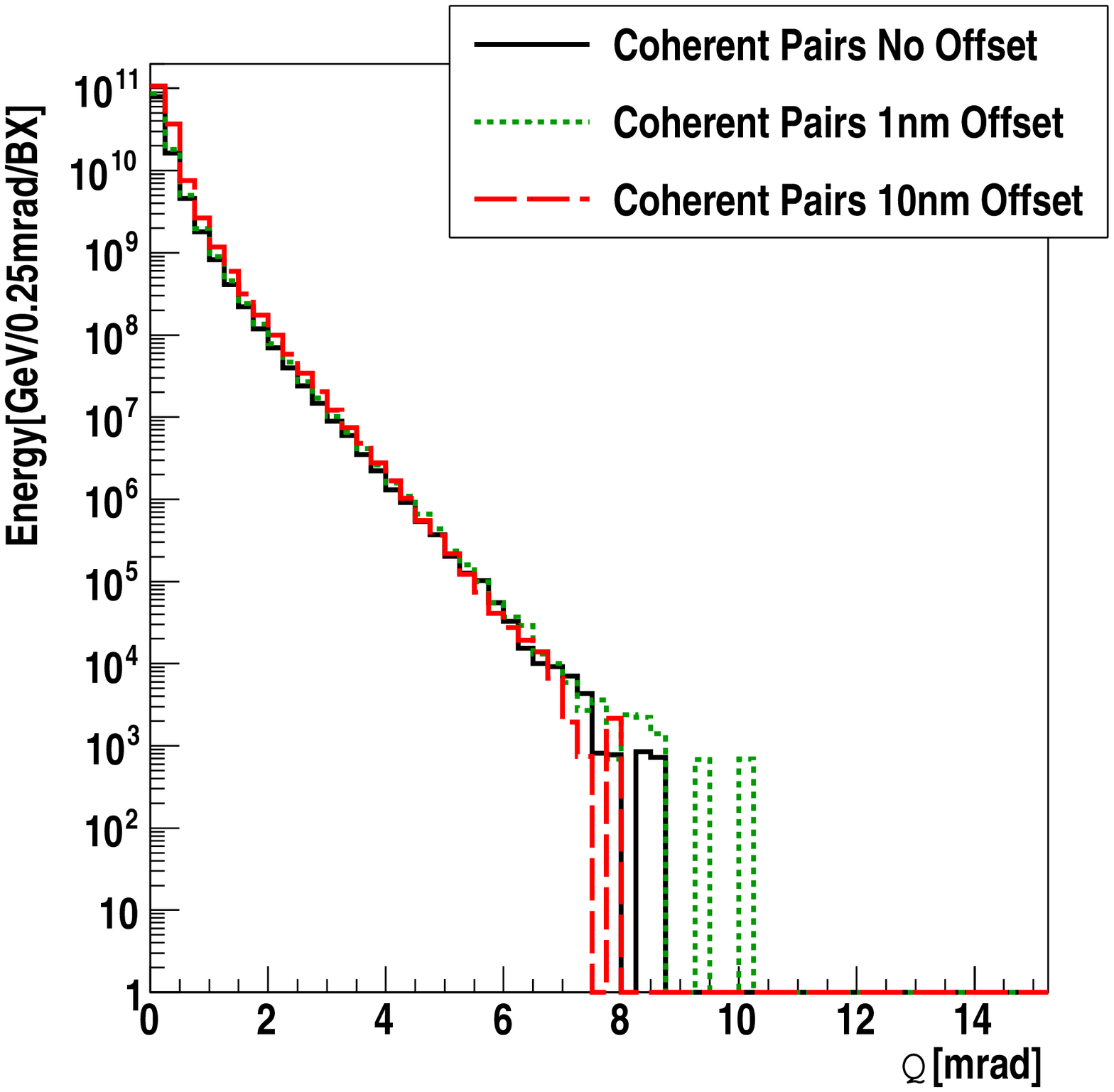}
  \end{tabular}
  \caption{Left: Number of incoherent and coherent pairs produced against vertical beam offset. The
    error bars are taking the large statistical weights of the coherent pairs
    into account. Right: Angular distribution of the energy from coherent pairs
    for different vertical offsets.}\label{fig:pairsvoff}
\end{figure}
The simulation of the beam-beam interaction is normally done for nominal
parameters, however, due to jitter in the accelerator, beam delivery system and
final focus quadrupole not all bunch collisions happen without offsets. To study
the impact of the vertical offset on the background several \textsc{GuineaPig}
runs with a vertical offset varied between 0 and 10~nm have been done. For an
offset of $\approx$2~nm, 50\% of the peak luminosity (with 1\% around the nominal
energy) is lost. Figure \ref{fig:pairsvoff} shows the number of pairs produced in
the coherent and incoherent process with respect to the offset. For small
offsets ($<0.3~\mathrm{nm}$) the number of both the incoherent and coherent
pairs changes less than 5\%. Since the luminosity drops rapidly with increasing
offsets, this is important because it implies that non-nominal collisions with
sufficient luminosity offer the same background environment as the nominal
collisions.

High energy electrons can be identified on top of the energy depositions in
BeamCal, if the fluctuations of the background is sufficiently small and well
known. It is therefore important to know, that the fluctuations in the
background from vertical offsets are small. Small vertical offsets should not
add significant fluctuations to the energy deposition in BeamCal.


For offsets above 10~nm, when only 2\% of the peak luminosity remains, the
coherent pairs could pose a problem for the mask, if their angular distribution
becomes wider.  Figure \ref{fig:pairsvoff} (right) shows the angular
distribution for different offsets, the cut-off fluctuates around 9~mrad so that
more statistics are needed for a clear answer.

\section{Simulation of the incoherent pair background}
\label{sec:bgsim}

\begin{figure}[tpb]
  \begin{tabular}{l r}
    \includegraphics[width=0.45\columnwidth]{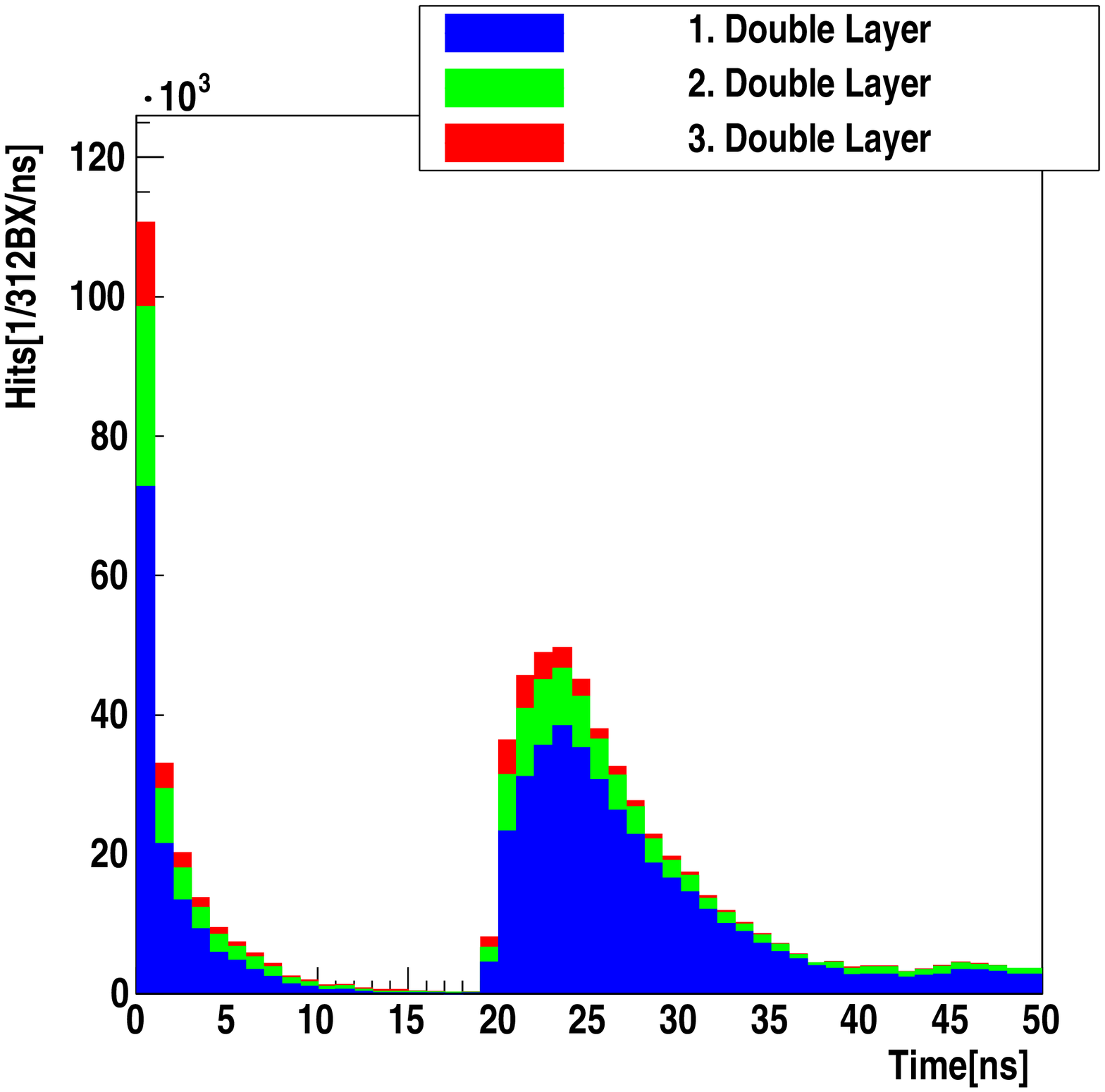} &
    \includegraphics[width=0.45\columnwidth]{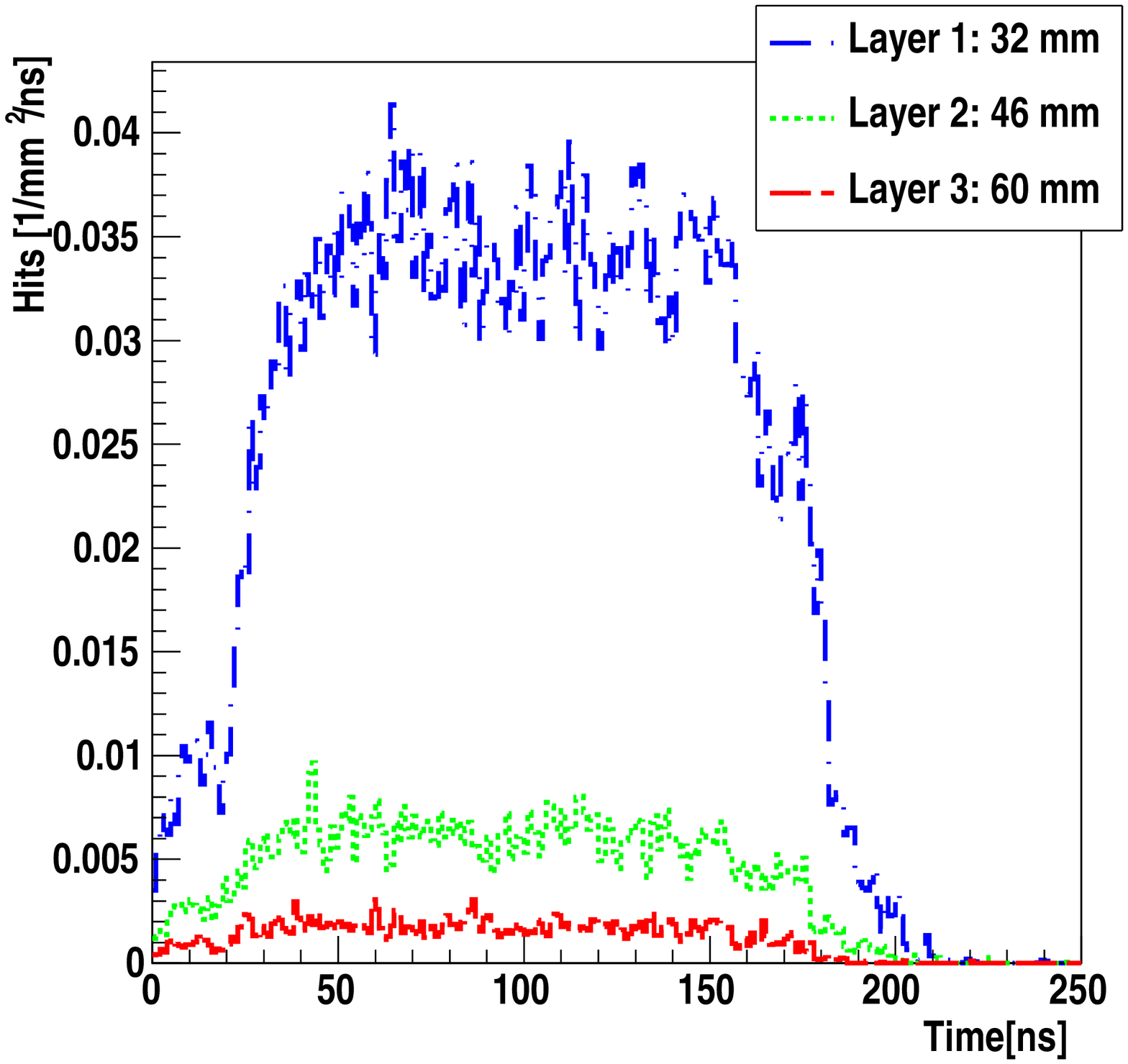}
  \end{tabular}
  \caption{Left: Time distribution of hits in the VXD for 312 BX; cumulative
    plot for the three double layers. Right: Average hit density in the VXD during a
    full bunch train.}\label{fig:hitsbt}
\end{figure}

\begin{wrapfigure}{O}{0.5\columnwidth}
  \centerline{\includegraphics[width=0.45\columnwidth]{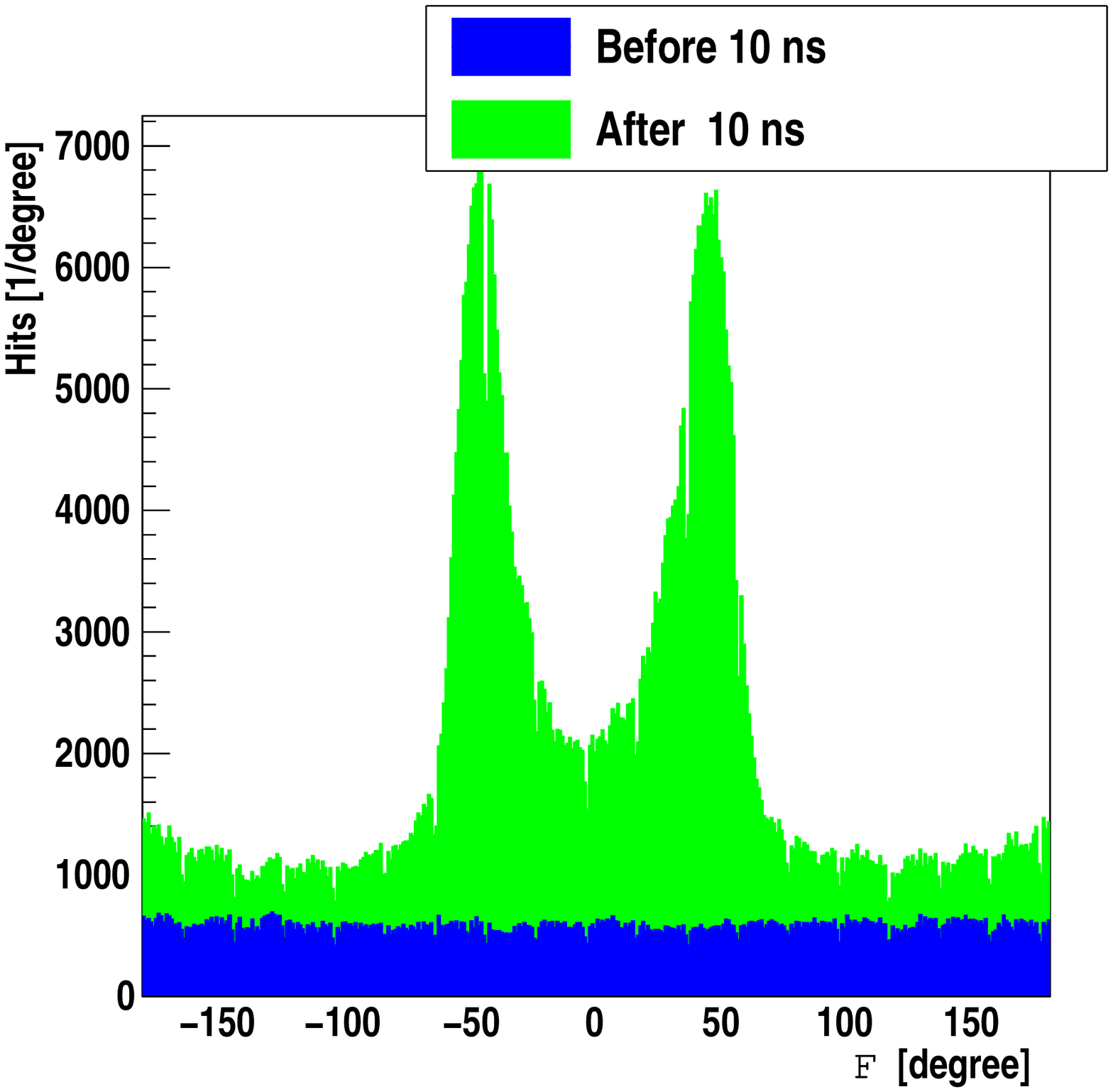}}
  \caption{Azimuthal distribution of hits for the first double layer of the VXD.}\label{fig:hitsphi}
\end{wrapfigure}

Since the coherent pairs are expected to leave the detector without touching any
material, only the incoherent pairs from the 312 bunch crossings are simulated
in the \textsc{Geant4}~\cite{Agostinelli2003} based full detector simulation
\textsc{Mokka}~\cite{Mora2002} with the CLIC detector and forward region. The
simulation have been performed with a \textsc{Geant4} range-cut of 0.005~mm and
the \texttt{QGSP\textunderscore{}BERT\textunderscore{}HP} physics list.

A hit in the VXD is counted if a particle deposits more than 3.4~keV in the
silicon of the sensor. No digitization is performed so far.  Figure
\ref{fig:hitsbt} (left) shows time distribution of hits in the VXD after the
bunch crossings. A clear separation between hits by particles coming directly
from the interaction point and by particles back-scattering from the forward
region can be seen. Direct hits can only be reduced by increasing the strength
of the solenoid field or the radius of the vertex
detector~\cite{schulte1999}. Back-scattering particles can be influenced by the
forward region design.  If the hit time is shifted according to the bunch
spacing within a train, the realistic hit densities can be estimated as shown on
the right in Figure \ref{fig:hitsbt}. The hit density is low during the first 20
nanoseconds, after which the back-scattered particles start to also hit the
vertex detector. If the vertex detector had to integrate over all the hits from
full bunch train (i.e. 156~ns) the hit density would correspond to
5.4~hits/mm$^2$, which might have a detrimental effect on the performance of the
pattern recognition and vertexing. To limit the impact a fast time-stamping in
the order of 5 to 20 nanoseconds is probably needed.

The hit density is comparable to the one found by simulations done for the
500~GeV ILC~\cite{wichmann2010}. In the first layer for a single bunch crossing
at CLIC or ILC they are the same with about 0.02~hits/mm$^2$, with a larger VXD
radius in the CLIC case.  Figure \ref{fig:hitsbt} shows that the hit density
strongly depends on the radius and decreases for larger radii.  Because no
Anti-DID field is used at the CLIC detector the distribution of hits is
inhomogeneous with respect to the azimuthal angle $\Phi$ (Figure
\ref{fig:hitsphi}). This inhomogeneity stems from back-scattering particles only
(labeled ``After 10~ns'') and occurs when low energy particles coming back from
BeamCal curl straight into the VXD. In this region the hit density is
considerably higher than the averages shown in Figure \ref{fig:hitsbt}
(right). The region with the highest number of hits will eventually limit the
detector life-time.

\section{Conclusions}
\label{sec:concs}

The incoherent pairs cause a large number of hits in the vertex detector, in the
first double layer this corresponds to an average hit density of 5.4~hits/mm$^2$
for a full bunch train. Two thirds of these are coming from back-scattering
particles. \textsc{GuineaPig} simulations show that the background environment
should not change for small vertical offsets of colliding bunches. The coherent
pairs also produced during the collisions should leave the detector region
without causing additional background hits, given that the aperture of the
beam-pipe is large enough, and this even with potentially large vertical beam
offsets at the interaction point.

\section{Acknowledgments}

I would like to express my gratitude to the \textsc{Mokka} developers,
especially Paulo Mora de Freitas, for the help to set up the
CLIC\textunderscore{}ILD simulation geometry. I would also like to thank Iftach
Sadeh and Adrian Vogel for valuable information regarding \textsc{Mokka} and the
members \textsc{FCal} collaboration for many discussions.

\begin{footnotesize}
\bibliography{lcd,beijing_proc_sailer_bib}

 \end{footnotesize}


\end{document}